\documentclass[groupedaddress,aps,pra,superscriptaddress,showpacs,twocolumn,prl]{revtex4}%
\usepackage{epsfig,dsfont,amssymb,amsmath,amsthm,amsfonts,amsbsy,mathrsfs}
\usepackage{graphicx, color}
\usepackage{epstopdf}
\usepackage{mathdots}
\usepackage{float}
\begin{document}

\title{Complementary Quantum Correlations among Multipartite Systems}

\author{Zhi-Xiang Jin}
\thanks{Corresponding author: jzxjinzhixiang@126.com}
\affiliation{School of Physics, University of Chinese Academy of Sciences, Yuquan Road 19A, Beijing 100049, China}
\author{Shao-Ming Fei}
\thanks{Corresponding author: feishm@mail.cnu.edu.cn}
\affiliation{School of Mathematical Sciences,  Capital Normal University,  Beijing 100048,  China}
\affiliation{Max-Planck-Institute for Mathematics in the Sciences, Leipzig 04103, Germany}
\author{Cong-Feng Qiao}
\thanks{Corresponding author: qiaocf@ucas.ac.cn}
\affiliation{School of Physics, University of Chinese Academy of Sciences, Yuquan Road 19A, Beijing 100049, China}
\affiliation{CAS Center for Excellence in Particle Physics, Beijing 100049, China\\ \vspace{7pt}}

\begin{abstract}
We study the monogamy and polygamy relations related to quantum correlations for multipartite quantum systems. General monogamy relations are presented for the $\alpha$th $(0\leq\alpha \leq\gamma, \gamma\geq2)$ power of quantum correlation, and general polygamy relations are given for the $\beta$th $(\beta\geq \delta, 0\leq\delta\leq1)$ power of quantum correlation. These monogamy and polygamy inequalities are complementary to the existing ones with different parameter regions of $\alpha$ and $\beta$. Applying these results to specific quantum correlations, the corresponding new classes of monogamy and polygamy relations are obtained, which include the existing ones as special cases. Detailed examples are given.
\end{abstract}
\maketitle

\section{INTRODUCTION}
Monogamy of entanglement is a representative feature of quantum physics, and therefore is a very important issue in the study of quantum information and quantum communication. It tells that the entanglement of a quantum system with one of the other ones limits its entanglements with the remaining systems. The monogamy property is highly related to quantum information processing tasks such as the security analysis of quantum key distribution \cite{MP}.	

The monogamy relation was first presented by Coffman, Kundu, and Wootters \cite{MK} for three-qubit states $\rho_{ABC}$, $\mathcal{E}(\rho_{A|BC})\geq \mathcal{E}(\rho_{AB}) +\mathcal{E}(\rho_{AC})$, where $\mathcal{E}$ is a bipartite entanglement measure, $\rho_{AB}$ and $\rho_{AC}$ are the reduced density matrices of $\rho_{ABC}$. Later on, the monogmay inequality was generalized to the case of multiqubit quantum system \cite{ZXN,JZX,jll}, high-dimensional quantum system \cite{012329} and general settings \cite{gy1,gy2,jin1,jin2}.
Polygamy of entanglement is characterized as, ${E_a}_{A|BC}\leq {E_a}_{AB} +{E_a}_{AC}$ for a tripartite quantum state $\rho_{ABC}$, where ${E_a}_{A|BC}$ is the assisted entanglement \cite{gg} between $A$ and $BC$. The first polygamy inequality was established in three-qubit systems by use of assisted entanglement \cite{gg}. It was later generalized to multiqubit systems \cite{jsb,jin3}. For the case of arbitrary-dimensional quantum systems, general polygamy inequalities of multipartite entanglement are also proposed in \cite{062328,295303,jsb,042332} by using entanglement of assistance.

Recently, monogamy and polygamy relations of multiqubit entanglement have been studied in terms of non-negative
power of entanglement measures and assisted entanglement measures. In \cite{JZX, jll,ZXN}, the authors have shown that the $x$th power of the entanglement of
formation ($x\geq\sqrt{2}$) and the concurrence ($x\geq2$) satisfy multiqubit monogamy inequalities. Monogamy relations for quantum steering have also been demonstrated in \cite{hqy,mko,jk1,jk2,jk3}.
Later, polygamy inequalities were proposed in terms of the
$\alpha$th ($0\leq\alpha\leq 1$) power of square of convex-roof extended negativity (SCREN) and the entanglement of assistance \cite{j012334, 042332}. In \cite{gy2}, the authors introduced the concept of polygamy relations without inequalities.

Whereas the monogamy of entanglement shows the restricted sharability of multipartite quantum entanglement, the distribution of entanglement in multipartite quantum systems was shown to have a dually monogamous property.
Based on concurrence of assistance \cite{qic}, the polygamy of entanglement provides a lower bound for the distribution of bipartite entanglement in a multipartite system \cite{jmp}.
Monogamy and polygamy of entanglement can restrict the possible correlations between the authorized users and the eavesdroppers, thus tightening the security bounds in quantum cryptography \cite{MP}.

However, the monogamy relations for the $\alpha$th $(0\leq\alpha\leq2)$ power and the polygamy relations for the $\beta$th $(\beta\geq1 )$ power of general quantum correlations are still not clear.
In this paper, we provide a class of monogamy and polygamy relations of the $\alpha$th $(0\leq\alpha\leq\gamma,\gamma\geq2)$ and the $\beta$th $(\beta\geq\delta,0\leq\delta\leq1)$ power for any quantum correlations. Application of the monogamy relations to quantum correlations like squared convex-roof extended negativity, entanglement of formation and concurrence give rise to tighter monogamy inequalities than the existing ones \cite{zhu} for some classes of quantum states. We take concurrence as an example to illustrate in detail.
Applying the general quantum correlations to specific quantum correlations, e.g. the concurrence of assistance, square of convex-roof extended negativity of assistance (SCRENoA), entanglement of assistance, the corresponding new class of polygamy relations are obtained, which are complementary to the existing ones \cite{jin3,062328,295303,jsb,042332} with different regions of parameter $\beta$.

\section{MONOGAMY RELATIONS for GENERAL quantum correlations}

Let $\mathcal{Q}$ be a measure of quantum correlation for bipartite systems. If a quantum measure $\mathcal{Q}$ satisfies \cite{ARA},
\begin{eqnarray}\label{q}
&&\mathcal{Q}(\rho_{A|B_1B_2,\cdots,B_{N-1}})\nonumber\\
&&\geq\mathcal{Q}(\rho_{AB_1})+\mathcal{Q}(\rho_{AB_2})+\cdots+\mathcal{Q}(\rho_{AB_{N-1}}),
\end{eqnarray}
we say it is monogamous, here $\rho_{AB_i}$, $i=1,...,N-1$, are the reduced density matrices of $\rho_{A|B_1B_2,\cdots,B_{N-1}}$. For simplicity, we denote $\mathcal{Q}(\rho_{AB_i})$ by $\mathcal{Q}_{AB_i}$, and $\mathcal{Q}(\rho_{A|B_1B_2,\cdots,B_{N-1}})$ by $\mathcal{Q}_{A|B_1B_2,\cdots,B_{N-1}}$. Some of the quantum measures have been shown to be monogamous \cite{AKE,SSS} for some classes states.
However, there exists some other measures which are known not satisfied the manogamy relations  \cite{GLGP, RPAK}.

In Ref. \cite{SPAU}, the authors have proved that there exists $\gamma\in R~(\gamma\geq2)$ such that, for arbitrary dimensional tripartite states, $\mathcal{Q}$ satisfies the following monogamy relation \cite{SPAU},
\begin{eqnarray}\label{aqq}
\mathcal{Q}^\gamma_{A|BC}\geq\mathcal{Q}^\gamma_{AB}+\mathcal{Q}^\gamma_{AC}.
\end{eqnarray}

In the following, we denote $\gamma$ the value such that $\mathcal{Q}$ satisfies the inequality (\ref{aqq}).  Using the inequality $(1+t)^x\geq 1+t^x$ for $x\geq1,~0\leq t\leq1$, it is easy to generalize the result (\ref{aqq}) to $N$-partite case,
\begin{eqnarray}\label{mq}
\mathcal{Q}^\gamma_{A|B_0B_1,\cdots,B_{N-1}}\geq \sum_{i=0}^{N-1}\mathcal{Q}_{AB_i}^\gamma.
\end{eqnarray}
First, we give a Lemma.

{[\bf Lemma 1]}. Suppose that $k$ is a real number satisfying $k\geq 1$. Then, for any real numbers $x$ and $t$, $0\leq x \leq 1$, $t\geq k$, we have $(1+t)^x\geq 1+\frac{(1+k)^x-1}{k^x}t^x$.

{\sf [Proof].} Let $f(x,y)=(1+y)^x-y^x$ with $0\leq x\leq 1,~0<y\leq \frac{1}{k}$. Then $\frac{\partial f}{\partial y}=x[(1+y)^{x-1}-y^{x-1}]\leq 0$. Therefore, $f(x,y)$ is a decreasing function of $y$, i.e., $f(x,y)\geq f(x,\frac{1}{k})=\frac{(1+k)^x-1}{k^x}$. Set $y=\frac{1}{t},~t\geq k$, we obtain $(1+t)^x\geq 1+\frac{(1+k)^x-1}{k^x}t^x$. $\Box$

{[\bf Theorem 1]}. Suppose that $k$ is a real number satisfying $k\geq 1$. Then, for any tripartite state $\rho_{ABC}$:

$(1)$ if $\mathcal{Q}^\gamma_{AC}\geq k\mathcal{Q}^\gamma_{AB}$, the quantum correlation measure $\mathcal{Q}$ satisfies
\begin{eqnarray}\label{th11}
\mathcal{Q}^\alpha_{A|BC}\geq\mathcal{Q}^\alpha_{AB}+\frac{(1+k)^\frac{\alpha}{\gamma}-1}{k^\frac{\alpha}{\gamma}}\mathcal{Q}^\alpha_{AC}
\end{eqnarray}
for $0\leq\alpha\leq \gamma$ and $\gamma\geq 2$.

$(2)$ if $\mathcal{Q}^\gamma_{AB}\geq k\mathcal{Q}^\gamma_{AC}$, the quantum correlation measure $\mathcal{Q}$ satisfies
\begin{eqnarray}\label{th12}
\mathcal{Q}^\alpha_{A|BC}\geq\mathcal{Q}^\alpha_{AC}+\frac{(1+k)^\frac{\alpha}{\gamma}-1}{k^\frac{\alpha}{\gamma}}\mathcal{Q}^\alpha_{AB}
\end{eqnarray}
for $0\leq\alpha\leq \gamma$ and $\gamma\geq 2$.

{\sf [Proof].} For arbitrary tripartite state $\rho_{ABC}$, one has \cite{SPAU},
$\mathcal{Q}^\gamma_{A|BC}\geq\mathcal{Q}^\gamma_{AB}+\mathcal{Q}^\gamma_{AC}$. If $\mathcal{Q}_{AB}~(\mathcal{Q}_{AC})=0$,
the inequality (\ref{th11}) or (\ref{th12}) are obvious. Therefore, assuming $\mathcal{Q}^\gamma_{AC}\geq k\mathcal{Q}^\gamma_{AB}>0$, we have
\begin{eqnarray}\label{pfth1}
\mathcal{Q}^{\gamma x}_{A|BC}&&\geq(\mathcal{Q}^\gamma_{AB}+\mathcal{Q}^\gamma_{AC})^x\nonumber\\
&&=\mathcal{Q}^{\gamma x}_{AB}\left(1+\frac{\mathcal{Q}^\gamma_{AC}}{\mathcal{Q}^\gamma_{AB}}\right)^x\nonumber\\
&&\geq\mathcal{Q}^{\gamma x}_{AB}\left(1+\frac{(1+k)^x-1}{k^x}\left(\frac{\mathcal{Q}^\gamma_{AC}}{\mathcal{Q}^\gamma_{AB}}\right)^x\right)\nonumber\\
&&=\mathcal{Q}^{\gamma x}_{AB}+\frac{(1+k)^x-1}{k^x}\mathcal{Q}^{\gamma x}_{AC},
\end{eqnarray}
where the second inequality is due to Lemma 1. Denote $\gamma x=\alpha$. Then $0\leq\alpha\leq \gamma$ as $0\leq x\leq1$, and we have the inequality (\ref{th11}). If $\mathcal{Q}_{AB}\geq k\mathcal{Q}_{AC}$, similarly we get (\ref{th12}).

{\sf [Remark 1].} We have presented a universal form of monogamy relations that are complementary to the existing ones in \cite{ZXN,JZX,jll} with different regions of the parameter $\alpha$ for any quantum correlations. Our general monogamy relations can be used to any quantum correlation measures like concurrence, negativity, entanglement of formation, and give rise to tighter monogamy relations than the existing ones in \cite{zhu} for some classes of quantum states. These monogamy relations can be also used to Tsallis-$q$ entanglement and Renyi-$q$ entanglement, which give new monogamy relations including the existing ones given in \cite{jll,jin3,slh} as special cases.

In the following, we take concurrence as an example to show the advantage of our conclusions.

Let $\mathds{H}_X$ denote a discrete finite-dimensional complex vector space associated with a quantum subsystem $X$.
For a bipartite pure state $|\psi\rangle_{AB}\in\mathds{H}_A\otimes \mathds{H}_B$, the concurrence is given by \cite{AU,PR,SA}, $C(|\psi\rangle_{AB})=\sqrt{{2\left[1-\mathrm{Tr}(\rho_A^2)\right]}}$,
where $\rho_A$ is the reduced density matrix obtained by tracing over the subsystem $B$, $\rho_A=\mathrm{Tr}_B(|\psi\rangle_{AB}\langle\psi|)$. The concurrence for a bipartite mixed state $\rho_{AB}$ is defined by the convex roof extension,
$C(\rho_{AB})=\min_{\{p_i,|\psi_i\rangle\}}\sum_ip_iC(|\psi_i\rangle)$,
where the minimum is taken over all possible decompositions of $\rho_{AB}=\sum\limits_{i}p_i|\psi_i\rangle\langle\psi_i|$, with $p_i\geq0$, $\sum\limits_{i}p_i=1$ and $|\psi_i\rangle\in \mathds{H}_A\otimes \mathds{H}_B$.

For an $N$-qubit state $\rho_{AB_1\cdots B_{N-1}}\in \mathds{H}_A\otimes \mathds{H}_{B_1}\otimes\cdots\otimes \mathds{H}_{B_{N-1}}$, if $C(\rho_{AB_i})\geq C(\rho_{A|B_{i+1}\cdots B_{N-1}})$ for $i=1, 2, \cdots, N-2$, $N\geq 4$, the concurrence satisfies \cite{jll},
\begin{eqnarray}\label{mo2}
&&C^\alpha(\rho_{A|B_1B_2\cdots B_{N-1}})\geq \sum_{j=1}^{N-1}(2^\frac{\alpha}{2}-1)^{j-1}C^\alpha(\rho_{AB_j}),
\end{eqnarray}
for $\alpha\geq2$.

For any $2\otimes2\otimes2^{N-2}$ tripartite mixed state $\rho_{ABC}$, if $C(\rho_{AC})\geq C(\rho_{AB})$, the concurrence satisfies \cite{zhu}
\begin{eqnarray}\label{mo3}
C^\alpha(\rho_{A|BC})\geq C^\alpha(\rho_{AB})+(2^\frac{\alpha}{\gamma}-1)C^\alpha(\rho_{AC})
\end{eqnarray}
for $0\leq\alpha\leq \gamma$ and $\gamma\geq 2$.

For concurrence, one has $\gamma\geq2$ \cite{TJ,YKM}. For convenience, we denote $C_{AB_i}=C(\rho_{AB_i})$ the concurrence of $\rho_{AB_i}$ and $C_{A|B_1,B_2\cdots,B_{N-1}}=C(\rho_{A|B_1\cdots B_{N-1}})$. Then, for the concurrence, we get the following conclusion by the similar method to the proof of Theorem 1.

{[\bf Corollary 1]}. Suppose that $k$ is a real number satisfying $k\geq 1$. Then, for any $2\otimes2\otimes2^{N-2}$ tripartite mixed state:

$(1)$ if $C^\gamma_{AC}\geq kC^\gamma_{AB}$, the concurrence satisfies
\begin{eqnarray}\label{co11}
C^\alpha_{A|BC}\geq C^\alpha_{AB}+\frac{(1+k)^\frac{\alpha}{\gamma}-1}{k^\frac{\alpha}{\gamma}}C^\alpha_{AC}
\end{eqnarray}
for $0\leq\alpha\leq \gamma$ and $\gamma\geq 2$.

$(2)$ if $C^\gamma_{AB}\geq kC^\gamma_{AC}$, the concurrence satisfies
\begin{eqnarray}\label{co12}
C^\alpha_{A|BC}\geq C^\alpha_{AC}+\frac{(1+k)^\frac{\alpha}{\gamma}-1}{k^\frac{\alpha}{\gamma}}C^\alpha_{AB}
\end{eqnarray}
for $0\leq\alpha\leq \gamma$ and $\gamma\geq 2$.

One can see that Corollary 1 reduces to the monogamy inequality (\ref{mo2}) for three-qubit states, if $k=1$, $\alpha=\gamma\geq2$, and reduces to the monogamy inequality (\ref{mo3}), if $k=1$. For $k>1$, the inequality (\ref{co11}) is tighter than the inequality (\ref{mo3}), as
$\frac{(1+k)^\frac{\alpha}{\gamma}-1}{k^\frac{\alpha}{\gamma}}\geq 2^\frac{\alpha}{\gamma}-1$, where the equality holds only for $\alpha=\gamma$.

{\it Example 1}. Let us consider the three-qubit state $|\psi\rangle$ in the generalized Schmidt decomposition form \cite{AA,XH},
\begin{eqnarray}\label{ex1}
|\psi\rangle&=&\lambda_0|000\rangle+\lambda_1e^{i{\varphi}}|100\rangle+\lambda_2|101\rangle \nonumber\\
&&+\lambda_3|110\rangle+\lambda_4|111\rangle,
\end{eqnarray}
where $\lambda_i\geq0,~i=0,1,2,3,4$ and $\sum\limits_{i=0}\limits^4\lambda_i^2=1.$
From the definition of concurrence, we have $C_{A|BC}=2\lambda_0\sqrt{{\lambda_2^2+\lambda_3^2+\lambda_4^2}}$, $C_{AB}=2\lambda_0\lambda_2$ and $C_{AC}=2\lambda_0\lambda_3$. Set $\lambda_{0}=\lambda_{3}=\frac{1}{2}$, $\lambda_{1}=\lambda_{2}=\lambda_{4}=\frac{\sqrt{6}}{6}$, $k=\frac{\sqrt{6}}{2}$, one has $C_{A|BC}=\frac{\sqrt{21}}{6}$, $C_{AB}=\frac{\sqrt{6}}{6}$, $C_{AC}=\frac{1}{2}$. Then $C_{A|BC}^{\alpha}=(\frac{\sqrt{21}}{6})^{\alpha}$, $C_{AB}^{\alpha}+(2^\frac{\alpha}{\gamma}-1)C_{AC}^{\alpha}=(\frac{\sqrt{6}}{6})^{\alpha}+(2^\frac{\alpha}{\gamma}-1)(\frac{1}{2})^\alpha$, $C_{AB}^{\alpha}+\frac{(1+k)^\frac{\alpha}{\gamma}-1}{k^\frac{\alpha}{\gamma}}C_{AC}^{\alpha}=(\frac{\sqrt{6}}{6})^{\alpha}+\frac{(1+k)^\frac{\alpha}{\gamma}-1}{k^\frac{\alpha}{\gamma}}(\frac{1}{2})^\alpha$.
One can see that our result is better
than the results in \cite{zhu} for $0\leq\alpha\leq 2$ and $\gamma\geq 2$, see Fig 1. To be more clear, set $z=\frac{(1+k)^\frac{\alpha}{\gamma}-1}{k^\frac{\alpha}{\gamma}}C_{AC}^{\alpha}-(2^\frac{\alpha}{\gamma}-1)C_{AC}^{\alpha}=\frac{(1+k)^\frac{\alpha}{\gamma}-1}{k^\frac{\alpha}{\gamma}}(\frac{1}{2})^\alpha-(2^\frac{\alpha}{\gamma}-1)(\frac{1}{2})^\alpha$, where $k=\frac{\sqrt{6}}{2}$, i.e., $z$ represents the difference of the concurrence between (\ref{mo3}) and (\ref{co11}) on the right side, see Fig. 2.
\begin{figure}
\centering
    \includegraphics[width=8cm]{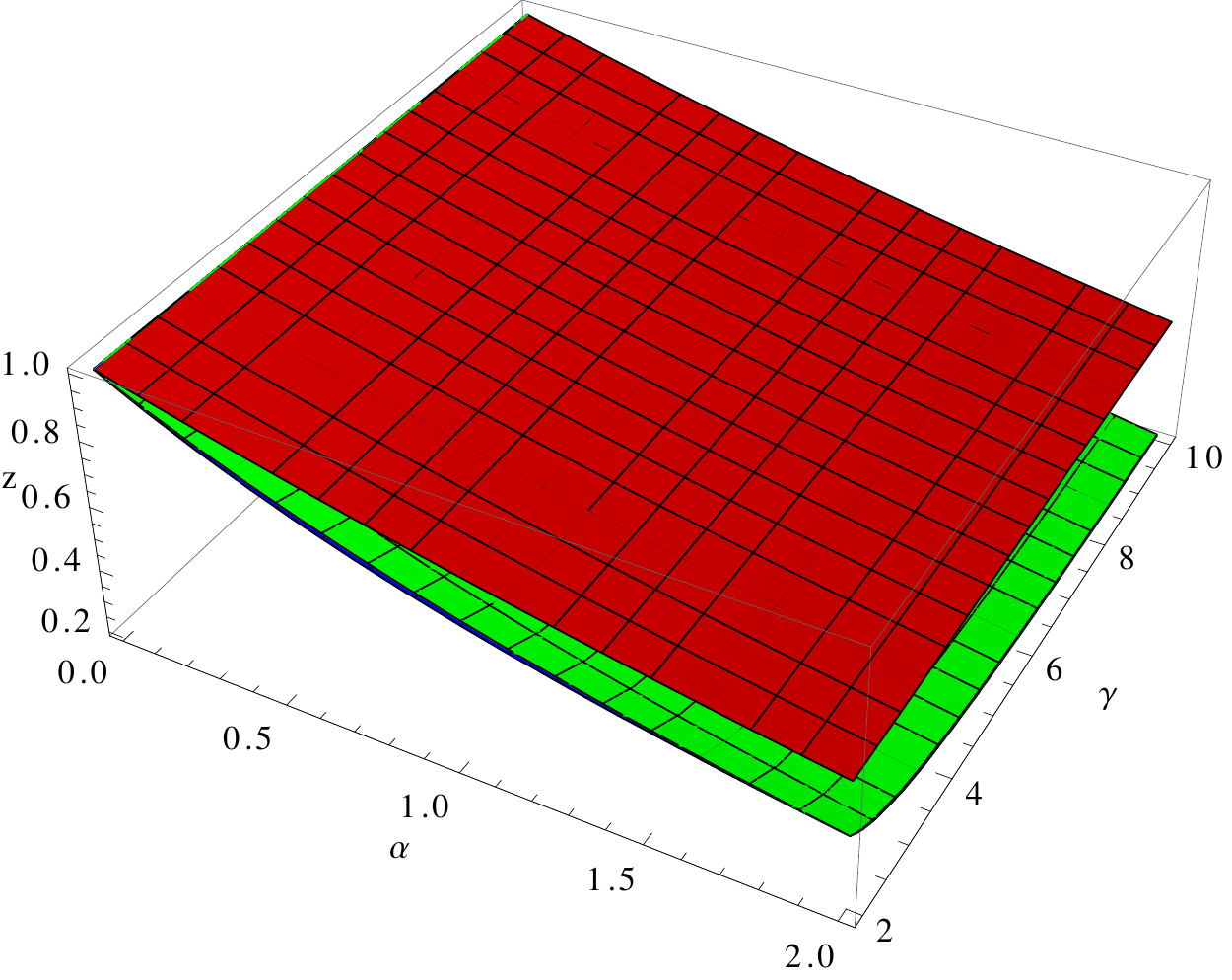}\\
  \caption{The axis $z$ is the concurrence of state $|\psi\rangle$ and its lower bounds, which are functions of $\alpha, \gamma$. The red surface represents the concurrence of the state $|\psi\rangle$, green surface represents the lower bound from our result, blue surface (just below the green one) represents the lower bound from the result in \cite{zhu}.}\label{2}
\end{figure}

\begin{figure}
\centering
    \includegraphics[width=8cm]{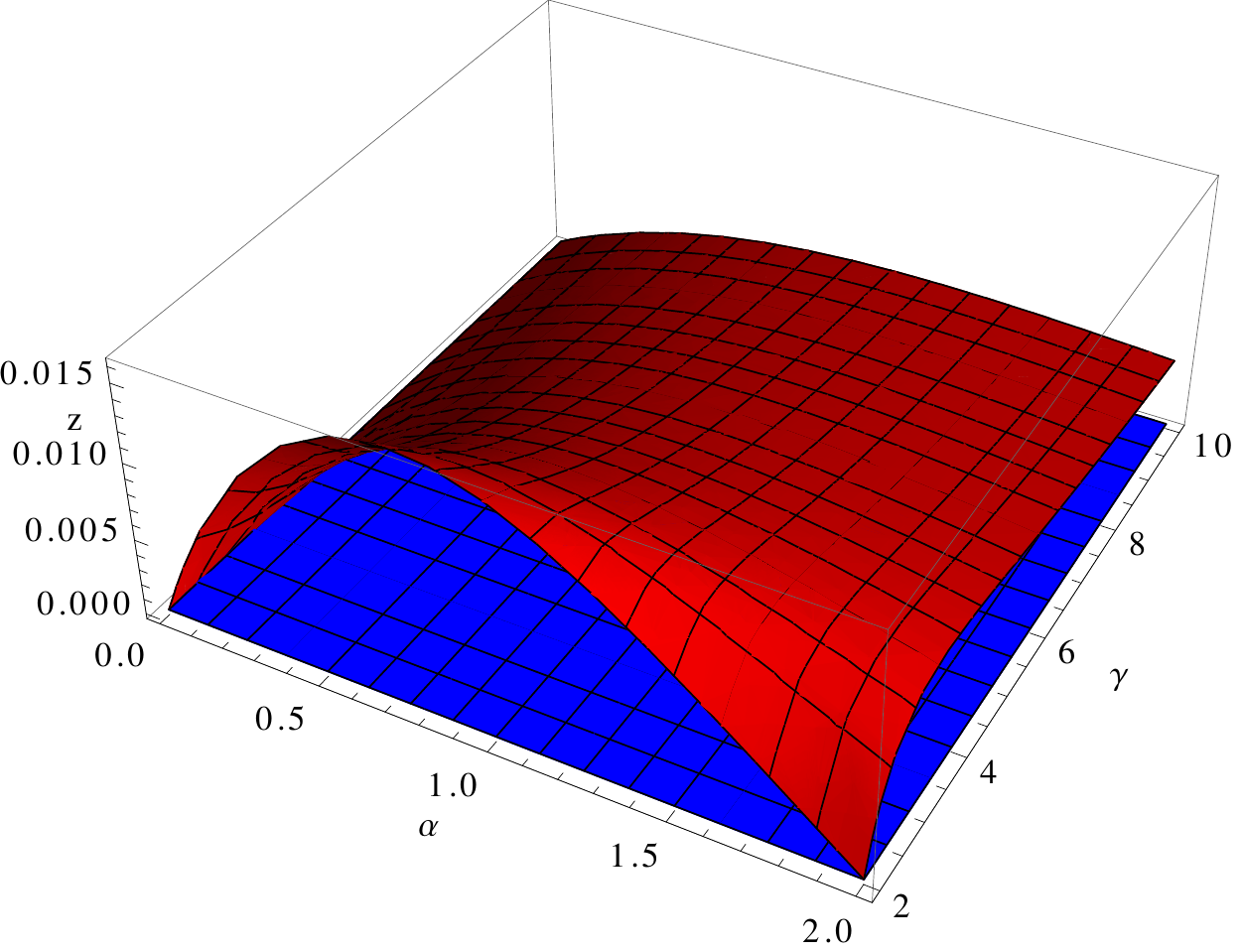}\\
  \caption{The red surface represents the difference of the concurrence between (\ref{mo3}) and (\ref{co11}) on the right side. The blue surface is zero plane of $z$.}\label{2}
\end{figure}

There are quantum correlation measures $\mathcal{Q}$ that themselves satisfy the usual monogamy relations, $\mathcal{Q}_{A|BC}\geq\mathcal{Q}_{AB}+\mathcal{Q}_{AC}$.
But generally, it is not the quantum correlation measure $\mathcal{Q}$ itself, but the
$\alpha$th power of the quantum correlation measure $\mathcal{Q}$ satisfies the monogamy inequality,
for instance, the $\alpha$th $(\alpha\geq2)$ power of concurrence and the $\alpha$th $(\alpha\geq\sqrt{2})$ power of the entanglement of formation \cite{ZXN}.
It is also the case for polygamy relations. Our Theorem 1 gives a general monogamy relation based on the $\alpha$th power of any quantum correlation measure.
For the entanglement measure concurrence in Corollary 1, as an example, one gets a tighter monogamy relation than the one in Ref. [25] for $0<\alpha\leq2$.
Monogamy relations characterize the distributions of quantum correlations in multipartite systems and play a crucial role in the security of quantum cryptography.
Tighter monogamy relations imply finer characterizations of the quantum correlation distributions, which tighten the security bounds in quantum cryptography.
The complementary monogamy relations may also help to investigate the efficiency of entanglement used in quantum cryptography \cite{30} and characterizations of the entanglement distributions.
These results may highlight future researches on quantum key distributions based on multipartite quantum entanglement distributions.

From the Example 1 above, one has that the relation $C_{A|BC}\geq C_{AB}+C_{AC}$ is not always satisfied for three-qubit states. In fact, the relation (\ref{aqq}), $C^\alpha_{A|BC}\geq C^\alpha_{AB}+C^\alpha_{AC}$, holds only for $\alpha\geq 2$ \cite{ZXN}. Nevertheless, the monogamy relations (\ref{co11}) holds for $0<\alpha<2$. In this sense the inequality (\ref{co11}) is complementary to (\ref{aqq}).
By using Theorem 1 repeatedly, we have the following theorem for multipartite quantum systems.

{[\bf Theorem 2]}. Suppose that $k$ is a real number satisfying $k\geq 1$. Then, for $N$-partite quantum state $\rho_{AB_1B_2\cdots B_{N-1}}$ such that
$k{\mathcal{Q}^\gamma_{AB_i}}\leq {\mathcal{Q}^\gamma_{A|B_{i+1}\cdots B_{N-1}}}$ for $i=1, 2, \cdots, m$, and
${\mathcal{Q}^\gamma_{AB_j}}\geq k{\mathcal{Q}^\gamma_{A|B_{j+1}\cdots B_{N-1}}}$ for $j=m+1,\cdots,N-2$,
$\forall$ $1\leq m\leq N-3$, $N\geq 4$, we have
\begin{eqnarray}\label{th1}
&&\mathcal{Q}^\alpha_{A|B_1B_2\cdots B_{N-1}}\geq \mathcal{Q}^\alpha_{AB_1}\nonumber \\
&&+\frac{(1+k)^\frac{\alpha}{\gamma}-1}{k^\frac{\alpha}{\gamma}} \mathcal{Q}^\alpha_{AB_2}+\cdots+\left(\frac{(1+k)^\frac{\alpha}{\gamma}-1}{k^\frac{\alpha}{\gamma}}\right)^{m-1}\mathcal{Q}^\alpha_{AB_m}\nonumber\\
&&+\left(\frac{(1+k)^\frac{\alpha}{\gamma}-1}{k^\frac{\alpha}{\gamma}}\right)^{m+1}(\mathcal{Q}^\alpha_{AB_{m+1}}+\cdots+\mathcal{Q}^\alpha_{AB_{N-2}}) \nonumber\\
&&+\left(\frac{(1+k)^\frac{\alpha}{\gamma}-1}{k^\frac{\alpha}{\gamma}}\right)^{m}\mathcal{Q}^\alpha_{AB_{N-1}}
\end{eqnarray}
for $0\leq\alpha\leq \gamma$ and $\gamma\geq 2$.

{\sf [Proof].} For convenience, we denote $l=\frac{(1+k)^\frac{\alpha}{\gamma}-1}{k^\frac{\alpha}{\gamma}}$. For $N$-partite quantum state $\rho_{AB_1B_2\cdots B_{N-1}}$, from the inequality (\ref{th11}), we have
\small
\begin{eqnarray}\label{pf11}
&&\mathcal{Q}^{\alpha}_{A|B_1B_2\cdots B_{N-1}}\nonumber\\
&&\geq  \mathcal{Q}^{\alpha}_{AB_1}+l\mathcal{Q}^{\alpha}_{A|B_2\cdots B_{N-1}}\nonumber\\
&&\geq \mathcal{Q}^{\alpha}_{AB_1}+l\mathcal{Q}^{\alpha}_{AB_2}
 +l^2\mathcal{Q}^{\alpha}_{A|B_3\cdots B_{N-1}}\nonumber\\
&& \geq \cdots\nonumber\\
&&\geq \mathcal{Q}^{\alpha}_{AB_1}+l\mathcal{Q}^{\alpha}_{AB_2}+\cdots+l^{m-1}\mathcal{Q}^{\alpha}_{AB_m}\nonumber\\
&&~~~~+l^m \mathcal{Q}^{\alpha}_{A|B_{m+1}\cdots B_{N-1}}.
\end{eqnarray}
\normalsize
Similarly, as ${\mathcal{Q}^\gamma_{AB_j}}\geq k{\mathcal{Q}^\gamma_{A|B_{j+1}\cdots B_{N-1}}}$ for $j=m+1,\cdots,N-2$, we get
\begin{eqnarray}\label{pf12}
&&\mathcal{Q}^{\alpha}_{A|B_{m+1}\cdots B_{N-1}} \nonumber\\
&&\geq l\mathcal{Q}^{\alpha}_{AB_{m+1}}+\mathcal{Q}^{\alpha}_{A|B_{m+2}\cdots B_{N-1}}\nonumber\\
&&\geq l(\mathcal{Q}^{\alpha}_{AB_{m+1}}+\cdots+\mathcal{Q}^{\alpha}_{AB_{N-2}})\nonumber\\
&&~~~~+\mathcal{Q}^{\alpha}_{AB_{N-1}}.
\end{eqnarray}
Combining (\ref{pf11}) and (\ref{pf12}), we have Theorem 2. $\Box$

We take concurrence as an example again to show the advantage of Theorem 2.

{[\bf Corollary 2]}. Suppose that $k\geq 1$. Then, for $N$-qubit quantum state $\rho_{AB_1B_2\cdots B_{N-1}}$ such that $k{C^\gamma_{AB_i}}\leq {C^\gamma_{A|B_{i+1}\cdots B_{N-1}}}$ for $i=1, 2, \cdots, m$, and
${C^\gamma_{AB_j}}\geq k{C^\gamma_{A|B_{j+1}\cdots B_{N-1}}}$ for $j=m+1,\cdots,N-2$,
$\forall$ $1\leq m\leq N-3$, $N\geq 4$, we have
\begin{eqnarray}\label{co2}
&&C^\alpha_{A|B_1B_2\cdots B_{N-1}}\geq \nonumber \\
&&C^\alpha_{AB_1}+lC^\alpha_{AB_2}+\cdots+l^{m-1}C^\alpha_{AB_m}\nonumber\\
&&+l^{m+1}(C^\alpha_{AB_{m+1}}+\cdots+C^\alpha_{AB_{N-2}}) \nonumber\\
&&+l^{m}C^\alpha_{AB_{N-1}}
\end{eqnarray}
for $0\leq\alpha\leq \gamma$, $\gamma\geq 2$ and  $l=\frac{(1+k)^\frac{\alpha}{\gamma}-1}{k^\frac{\alpha}{\gamma}}$.

For an $N$-qubit quantum state $\rho_{AB_1B_2\cdots B_{N-1}}$, it has been shown in \cite{ZXN} that the
$\alpha$th concurrence $C^\alpha$ ($0<\alpha<2$) does not satisfy monogamy inequalities like $C^\alpha(|\psi\rangle_{AB_1B_2\cdots B_{N-1}})\geq\sum_{i=1}^{N-1}C^\alpha(\rho_{AB_i})$. Theorem 2 gives a general monogamy
inequality satisfied by the $\alpha$th power of quantum correlation for the case of $0<\alpha<\gamma$ and $\gamma\geq 1$. Specifically, using the concurrence as an example, we obtain the monogamy inequality satisfied by $\alpha$th power of concurrence $C^\alpha$ for the case of $0<\alpha<2$, which was absent in \cite{jll}. Furthermore, inequality (\ref{co2}) in Corollary 2 reduces to the monogamy inequality (\ref{mo2}) if $k=1$ and $\alpha=\gamma\geq2$, and to the main result in \cite{zhu} for $k=1$. For $k>1$, the inequality (\ref{co2}) is tighter than the result in \cite{zhu}, since
$\frac{(1+k)^\frac{\alpha}{\gamma}-1}{k^\frac{\alpha}{\gamma}}\geq 2^\frac{\alpha}{\gamma}-1 (0<\alpha\leq\gamma)$, in which the equality holds only for $\alpha=\gamma$. The monogamy relations can also be applied to other specific quantum correlations, and similar new results can be obtained.

\section{polygamy RELATIONS for GENERAL quantum correlations}

 In \cite{jinzx}, the authors proved that for arbitrary dimensional tripartite states, there exists $\delta\in R~(0\leq\delta\leq1)$ such that a quantum correlation measure $\mathcal{Q}$ satisfies the following polygamy relation,
\begin{eqnarray}\label{aq}
\mathcal{Q}^\delta_{A|BC}\leq\mathcal{Q}^\delta_{AB}+\mathcal{Q}^\delta_{AC}.
\end{eqnarray}
In the follwing, we introduce the polygamy relations of the $\beta$th $(\beta\geq\delta)$ power for general quantum correlations, which is still not clear up to now. We first give a Lemma.

{[\bf Lemma 2]}. Suppose that $k$ is a real number satisfying $k\geq 1$. Then, for any real numbers $x$ and $t$, $x \geq 1$, $t\geq k$, we have $(1+t)^x\leq 1+\frac{(1+k)^x-1}{k^x}t^x$.

{\sf [Proof].} Let $f(x,y)=(1+y)^x-y^x$ with $x\geq 1,~0<y\leq \frac{1}{k}$. Then $\frac{\partial f}{\partial y}=x[(1+y)^{x-1}-y^{x-1}]\geq 0$. Therefore, $f(x,y)$ is an increasing function of $y$, i.e., $f(x,y)\leq f(x,\frac{1}{k})=\frac{(1+k)^x-1}{k^x}$. Set $y=\frac{1}{t},~t\geq k$, we obtain $(1+t)^x\leq 1+\frac{(1+k)^x-1}{k^x}t^x$. $\Box$

Using the similar method to the proof of Theorem 1 and the Lemma 2, we have

{[\bf Theorem 3]}. Suppose that $k$ is a real number satisfying $k\geq 1$. Then, for any tripartite state $\rho_{ABC}\in H_A\otimes H_B\otimes H_C$:

$(1)$ if $\mathcal{Q}^\delta_{AC}\geq k\mathcal{Q}^\delta_{AB}$, the quantum correlation measure $\mathcal{Q}$ satisfies
\begin{eqnarray}\label{th31}
\mathcal{Q}^\beta_{A|BC}\leq\mathcal{Q}^\beta_{AB}+\frac{(1+k)^\frac{\beta}{\delta}-1}{k^\frac{\beta}{\delta}}\mathcal{Q}^\beta_{AC}
\end{eqnarray}
for $\beta\geq \delta$ and $0\leq\delta\leq1$.

$(2)$ if $\mathcal{Q}^\delta_{AB}\geq k\mathcal{Q}^\delta_{AC}$, the quantum correlation measure $\mathcal{Q}$ satisfies
\begin{eqnarray}\label{th32}
\mathcal{Q}^\beta_{A|BC}\leq\mathcal{Q}^\beta_{AC}+\frac{(1+k)^\frac{\beta}{\delta}-1}{k^\frac{\beta}{\delta}}\mathcal{Q}^\beta_{AB}
\end{eqnarray}
for $\beta\geq \delta$ and $0\leq\delta\leq1$.

By using Theorem 3 repeatedly, with the similar method to the proof of Theorem 2, we have the following theorem for multipartite quantum systems.

{[\bf Theorem 4]}. Suppose that $k$ is a real number satisfying $k\geq 1$. Then, for $N$-partite quantum state $\rho_{AB_1B_2\cdots B_{N-1}}$ such that
$k{\mathcal{Q}^\delta_{AB_i}}\leq {\mathcal{Q}^\delta_{A|B_{i+1}\cdots B_{N-1}}}$ for $i=1, 2, \cdots, m$, and
${\mathcal{Q}^\delta_{AB_j}}\geq k{\mathcal{Q}^\delta_{A|B_{j+1}\cdots B_{N-1}}}$ for $j=m+1,\cdots,N-2$,
$\forall$ $1\leq m\leq N-3$, $N\geq 4$, we have
\begin{eqnarray}\label{th4}
&&\mathcal{Q}^\beta_{A|B_1B_2\cdots B_{N-1}}\leq \mathcal{Q}^\beta_{AB_1}\nonumber \\
&&+\frac{(1+k)^\frac{\beta}{\delta}-1}{k^\frac{\beta}{\delta}} \mathcal{Q}^\beta_{AB_2}+\cdots+\left(\frac{(1+k)^\frac{\beta}{\delta}-1}{k^\frac{\beta}{\delta}}\right)^{m-1}\mathcal{Q}^\beta_{AB_m}\nonumber\\
&&+\left(\frac{(1+k)^\frac{\beta}{\delta}-1}{k^\frac{\beta}{\delta}}\right)^{m+1}(\mathcal{Q}^\beta_{AB_{m+1}}+\cdots+\mathcal{Q}^\beta_{AB_{N-2}}) \nonumber\\
&&+\left(\frac{(1+k)^\frac{\beta}{\delta}-1}{k^\frac{\beta}{\delta}}\right)^{m}\mathcal{Q}^\beta_{AB_{N-1}}
\end{eqnarray}
for $\beta\geq \delta$ and $0\leq\delta\leq1$.

{\sf [Remark 2].} We present a universal form of polygamy relations that are complementary to the existing ones in \cite{jin3,062328,295303,jsb,042332} with different regions of parameter $\beta$ for any quantum correlations. Our general monogamy relations can be used to any quantum correlation measures like concurrence of assistance, square of convex-roof extended negativity of assistance (SCRENoA), entanglement of assistance. Corresponding new class of polygamy relations can be obtained. In the following, we take SCRENoA as an example.

Given a bipartite state $\rho_{AB}$ in $H_A\otimes H_B$, the negativity is defined by \cite{GRF},
$N(\rho_{AB})=(||\rho_{AB}^{T_A}||-1)/2$,
where $\rho_{AB}^{T_A}$ is the partially transposed $\rho_{AB}$ with respect to the subsystem $A$, $||X||$ denotes the trace norm of $X$, i.e $||X||=\mathrm{Tr}\sqrt{XX^\dag}$.
 For the purpose of discussion, we use the following definition of negativity, $ N(\rho_{AB})=||\rho_{AB}^{T_A}||-1$.
For any bipartite pure state $|\psi\rangle_{AB}$, the negativity $ N(\rho_{AB})$ is given by
$N(|\psi\rangle_{AB})=2\sum_{i<j}\sqrt{\lambda_i\lambda_j}=(\mathrm{Tr}\sqrt{\rho_A})^2-1$,
where $\lambda_i$ are the eigenvalues for the reduced density matrix $\rho_A$ of $|\psi\rangle_{AB}$. For a mixed state $\rho_{AB}$, the square of convex-roof extended negativity (SCREN) is defined by
 $N_{sc}(\rho_{AB})=[\mathrm{min}\sum_ip_iN(|\psi_i\rangle_{AB})]^2$,
where the minimum is taken over all possible pure state decompositions $\{p_i,~|\psi_i\rangle_{AB}\}$ of $\rho_{AB}$. The SCRENoA is defined by $N_{sc}^a(\rho_{AB})=[\mathrm{max}\sum_ip_iN(|\psi_i\rangle_{AB})]^2$, where the maximum is taken over all possible pure state decompositions $\{p_i,~|\psi_i\rangle_{AB}\}$ of $\rho_{AB}$. For convenience, we denote ${N_a}_{AB_i}=N_{sc}^a(\rho_{AB_i})$ the SCRENoA of $\rho_{AB_i}$ and ${N_a}_{AB_0,B_1\cdots,B_{N-1}}=N^a_{sc}(|\psi\rangle_{AB_0\cdots B_{N-1}})$.

In \cite{j012334} it has been shown that
${N_a}_{A|B_1\cdots B_{N-1}}\leq \sum_{j=1}^{N-1}{N_a}_{AB_j}.$
It is further improved that for $0\leq\beta\leq1$ \cite{jin3},
\begin{eqnarray}\label{n5}
{N_a}^\beta_{A|B_1\cdots B_{N-1}}\leq\sum_{j=1}^{N-1} (2^\beta-1)^j{N_a}^\beta_{AB_j}.
\end{eqnarray}

For SCRENoA, one has $0\leq\delta\leq 1$, using the Theorem 3 directly, we have

{[\bf Corollary 3]}. Suppose that $k$ is a real number satisfying $k\geq 1$. Then, for any $2\otimes2\otimes2^{N-2}$ tripartite mixed state:

$(1)$ if ${N_a}^\delta_{AC}\geq k{N_a}^\delta_{AB}$, the SCRENoA satisfies
\begin{eqnarray}\label{co31}
{N_a}^\beta_{A|BC}\leq{N_a}^\beta_{AB}+\frac{(1+k)^\frac{\beta}{\delta}-1}{k^\frac{\beta}{\delta}}{N_a}^\beta_{AC}
\end{eqnarray}
for $\beta\geq \delta$ and $0\leq\delta\leq1$.

$(2)$ if ${N_a}^\delta_{AB}\geq k{N_a}^\delta_{AC}$, the SCRENoA satisfies
\begin{eqnarray}\label{th32}
{N_a}^\beta_{A|BC}\leq{N_a}^\beta_{AC}+\frac{(1+k)^\frac{\beta}{\delta}-1}{k^\frac{\beta}{\delta}}{N_a}^\beta_{AB}
\end{eqnarray}
for $\beta\geq \delta$ and $0\leq\delta\leq1$.

For multiqubit quantum state $\rho_{AB_1B_2\cdots B_{N-1}}$, we have

{[\bf Corollary 4]}. Suppose that $k$ is a real number satisfying $k\geq 1$. Then, for $N$-qubit quantum state $\rho_{AB_1B_2\cdots B_{N-1}}$ such that
$k{{N_a}^\delta_{AB_i}}\leq {{N_a}^\delta_{A|B_{i+1}\cdots B_{N-1}}}$ for $i=1, 2, \cdots, m$, and
${{N_a}^\delta_{AB_j}}\geq k{{N_a}^\delta_{A|B_{j+1}\cdots B_{N-1}}}$ for $j=m+1,\cdots,N-2$,
$\forall$ $1\leq m\leq N-3$, $N\geq 4$, we have
\begin{eqnarray}\label{co4}
&&{N_a}^\beta_{A|B_1B_2\cdots B_{N-1}}\leq {N_a}^\beta_{AB_1}\nonumber \\
&&+\frac{(1+k)^\frac{\beta}{\delta}-1}{k^\frac{\beta}{\delta}} {N_a}^\beta_{AB_2}+\cdots+\left(\frac{(1+k)^\frac{\beta}{\delta}-1}{k^\frac{\beta}{\delta}}\right)^{m-1}{N_a}^\beta_{AB_m}\nonumber\\
&&+\left(\frac{(1+k)^\frac{\beta}{\delta}-1}{k^\frac{\beta}{\delta}}\right)^{m+1}({N_a}^\beta_{AB_{m+1}}+\cdots+{N_a}^\beta_{AB_{N-2}}) \nonumber\\
&&+\left(\frac{(1+k)^\frac{\beta}{\delta}-1}{k^\frac{\beta}{\delta}}\right)^{m}{N_a}^\beta_{AB_{N-1}}
\end{eqnarray}
for $\beta\geq \delta$ and $0\leq\delta\leq1$.

One can see that Corollary 4 reduces to the monogamy inequality (\ref{n5}), if $k=1$ and $0\leq\beta=\delta\leq1$.
\begin{figure}
\centering
    \includegraphics[width=8cm]{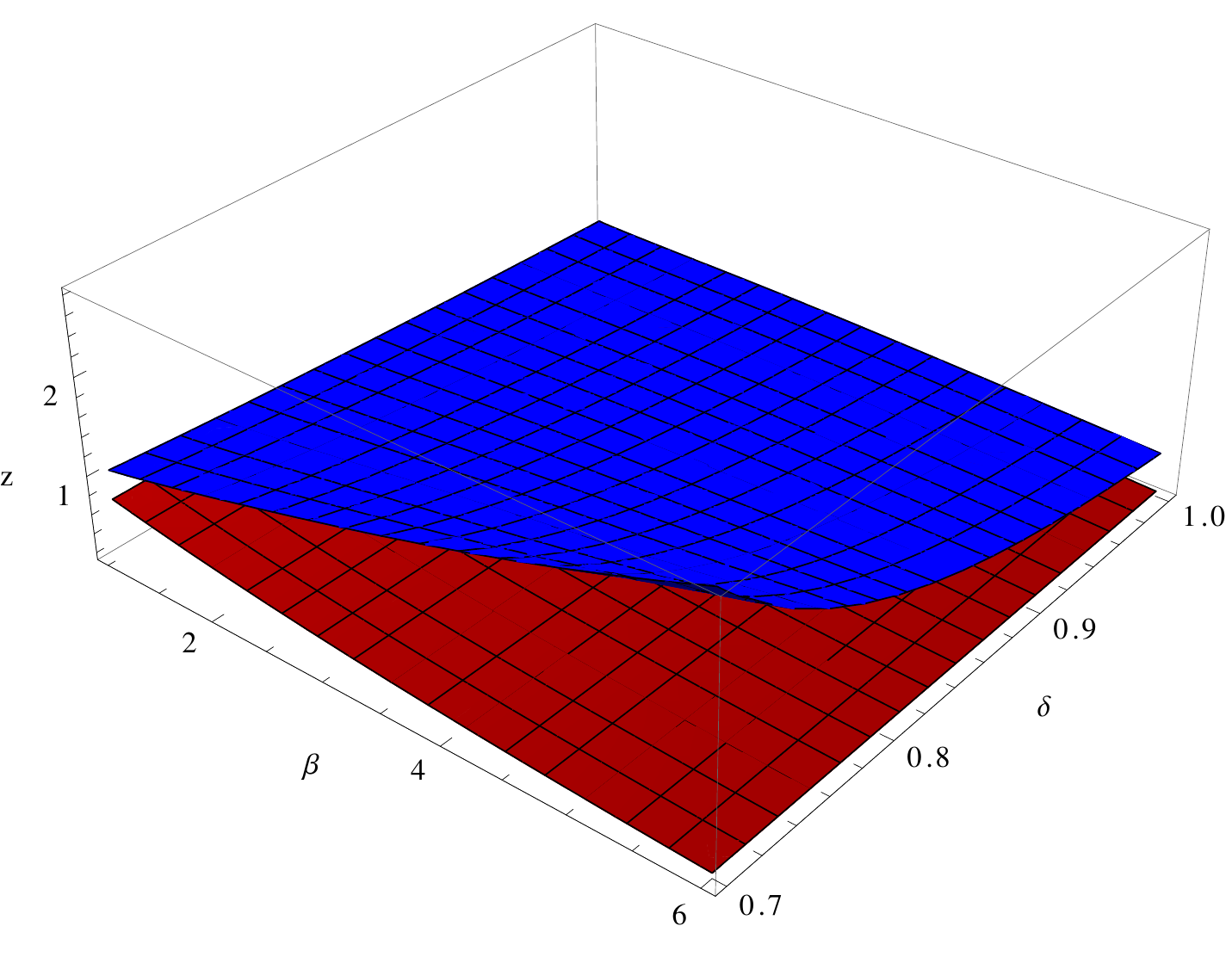}\\
  \caption{The axis $z$ is the SCRENoA of the state $|W\rangle_{AB_1B_2}$ and its upper bound, which are functions of $\beta, \delta$. The red surface represents the SCRENoA of the state $|W\rangle_{AB_1B_2}$, blue surface represents the upper bound from inequality (\ref{co4}).}\label{2}
\end{figure}
{\it Example 3}. Let us consider the three-qubit generlized $W$-class states,
\begin{eqnarray}\label{W}
|W\rangle_{AB_1B_2}=\frac{1}{2}(|100\rangle+|010\rangle)+\frac{\sqrt{2}}{2}|001\rangle.
\end{eqnarray}
We have ${N_a}_{A|B_1B_2}=\frac{3}{4}$, ${N_a}_{AB_1}=\frac{1}{4},~{N_a}_{AB_2}=\frac{1}{2}$. Then ${N^\beta_a}_{A|B_1B_2}=(\frac{3}{4})^\beta$, ${N^\beta_a}_{AB_1}+\frac{(1+k)^\frac{\beta}{\delta}-1}{k^\frac{\beta}{\delta}}{N^\beta_a}_{AB_2}=(\frac{1}{4})^\beta+\frac{(1+k)^\frac{\beta}{\delta}-1}{k^\frac{\beta}{\delta}}(\frac{1}{2})^\beta$, see Fig. 3.

\section{conclusion}
We investigate in this work the monogamy and polygamy relations related to quantum correlations for multipartite quantum systems. General monogamy relations are obtained for the $\alpha$th $(0\leq\alpha\leq\gamma, \gamma\geq2)$ power of quantum correlations, as well as the polygamy relations of the $\beta$th $(\beta \geq\delta,0\leq\delta\leq1 )$ power. These novel monogamy and polygamy inequalities are complementary to the existing ones with different regions of $\alpha$ and $\beta$. Applying the general quantum correlations to specific quantum correlations, the corresponding new class of monogamy and polygamy relations are obtained, which include the existing ones as special cases. To be noted that our approach is applicable to the study of the monogamy and polygamy relations of high dimensional quantum system.

\bigskip

\noindent{\bf Acknowledgments}\, \,
We thank anonymous reviewers for their suggestions in improving the manuscript.
This work was supported in part by the National Natural Science Foundation of China(NSFC) under Grants 11847209; 11675113 and 11635009; the Key Project of Beijing Municipal Commission of Education (Grant No.
KZ201810028042); the Beijing Natural Science Foundation (Z190005);
the Ministry of Science and Technology of the Peoples' Republic of China (2015CB856703); the Strategic Priority Research Program of the Chinese Academy of Sciences, Grant No. XDB23030100 and the China Postdoctoral Science Foundation funded project.

\end{document}